\documentclass[
    reprint,
    preprintnumbers,
    superscriptaddress,
    nofootinbib,
    amsmath,amssymb,
    aps,
    prl,
    floatfix,
    longbibliography]{revtex4-2}
    
    \usepackage{graphicx}
    \usepackage{subfigure}
    \usepackage{multirow}
    \usepackage[usenames,dvipsnames]{xcolor} 
    \usepackage{mathrsfs} 

    \usepackage{pgfplots}
    \usepackage[utf8]{inputenc}
    \usepackage{color}
    \usepackage{hyperref}
    \usepackage[normalem]{ulem} 
    \usepackage{physics}
    \usepackage{enumitem}
    \usepackage{comment}
    \usepackage{bm}
    \usepackage{aas_macros}
    \usepackage[capitalise]{cleveref}
    \usepackage{multirow}
    \usepackage{mathrsfs}
    \usepackage{orcidlink}

\hypersetup{
  breaklinks = true,
  colorlinks   = true, 
  urlcolor     = Blue, 
  linkcolor    = Blue, 
  citecolor   = Blue 
}

\newcommand{\set}[1]{\left\{#1\right\}}

\def\actaa{\ref@jnl{Acta Astron.}}      
\def\apjl{\textrm{ApJ}}                
\def\apjs{\textrm{ApJS}}               
\def\ao{\textrm{Appl.~Opt.}}           
\def\aap{\textrm{A\&A}}                

\usepackage{float}
\usepackage[normalem]{ulem}

\newcommand{\md}{{\rm d}}

\newcommand{\oxford}{Department of Physics, University of Oxford, Parks Road, Oxford, OX1 3PU, United Kingdom}

\newcommand{\splitatcommas}[1]{%
  \begingroup
  \begingroup\lccode`~=`, \lowercase{\endgroup
    \edef~{\mathchar\the\mathcode`, \penalty0 \noexpand\hspace{0pt plus 1em}}%
  }\mathcode`,="8000 #1%
  \endgroup
}

\pgfplotsset{compat=1.18}

\begin{document}

\title{Apparent Dark-Energy Evolution from Cosmic Inhomogeneities}

\author{Yonadav Barry Ginat\,\orcidlink{0000-0003-1992-1910}}
\email{yb.ginat@physics.ox.ac.uk}
\affiliation{\oxford}%
\affiliation{New College, Holywell Street, Oxford, OX1 3BN, United Kingdom}
\author{Pedro G.~Ferreira\,\orcidlink{0000-0002-3021-2851}}
\email{pedro.ferreira@physics.ox.ac.uk}
\affiliation{\oxford}

\begin{abstract}
	A mildly inhomogeneous universe with a cosmological constant may look like it contains evolving dark energy. We show that could be the case by modelling the inhomogeneities and their effects in three different ways: as clumped matter surrounded by voids, as back-reaction of small-scale structure on the overall expansion of the Universe, and, finally, as a large-scale curvature inhomogeneity. In all of these cases, the propagation of light is affected, and differs from that in a homogeneous and isotropic universe. The net result is that cosmological observables, such as angular diameter and luminosity distances, become distorted. We find, in all three models, that the inclusion of these effects pushes the distance--redshift relation towards closer agreement with recent data from both supernov{\ae} Ia from the Dark Energy Survey, and from baryon acoustic oscillations from the Dark Energy Spectroscopic Instrument. The amount of inhomogeneity in these models might not be enough to explain the entirety of the deviation from a cosmological constant, but is found to be of a similar order of magnitude, hinting that these data may be consistent with a universe dominated by a cosmological constant.
\end{abstract}

\maketitle



\paragraph{Introduction.}
The most recent measurements of the Universe's expansion rate suggest that, at late times, it may be dominated by evolving dark energy \cite{DES:2024jxu,DESI:2024mwx,DESI:2025zgx}. This would be striking, since the current favoured model assumes a time-independent cosmological constant, $\Lambda$. These results have prompted renewed interest in dark-energy scenarios with so-called {\it phantom} behaviour, in which the effective energy density grows and later decays \cite{Wolf:2023uno,Wolf:2024eph,Shlivko:2024llw, Tada:2024znt, Payeur:2024dnq, Gialamas:2025pwv, Lodha:2025qbg,Luu:2025fgw, Mishra:2025goj, Bhattacharya:2024hep, Wang:2024hwd, Wolf:2024stt,Wolf:2025jed, Ye:2024ywg, Chudaykin:2024gol, Ye:2025ulq,Pan:2025psn,Goldstein:2025epp,Cai:2025mas,Adam:2025kve,Khoury:2025txd,Chakraborty:2025syu,vanderWesthuizen:2025rip,Shah:2025ayl,Braglia:2025gdo,Chen:2025ywv,Andriot:2025los,Li:2024qso}.

It may be premature to infer the existence of evolving dark energy. Constraints from type Ia supernov{\ae} (SNe) depend strongly on a very low-redshift subset \cite{DESI:2025zgx}, though surveys are under way to regenerate or replace it \cite{2025A&A...694A...1R,2025arXiv251221903M}; moreover, re-calibration of the Dark Energy Survey SNe I${\rm a}$ photometric redshifts has weakened the evidence \cite{DES-Dovekie2026}. Constraints from baryon acoustic oscillations (BAO) vary with the data-set, and an unexplained inconsistency between those of the Dark Energy Spectroscopic Instrument (DESI) and the Sloan Digital Sky Survey (SDSS) may indicate residual systematics \cite{Efstathiou:2024dvn,DESI:2025zgx}. Finally, cosmic-microwave-background (CMB) constraints are also sensitive to the data-set and to parameter assumptions, e.g.~the optical depth \cite{Sailer:2025lxj} or spatial curvature \cite{Chen:2025mlf}, which can further weaken the case for evolving dark energy.

In this paper we will muddy the waters even further. While the Universe can be fairly described as homogeneous and isotropic on large scales, we know that is not exactly the case. One expects that deviations from homogeneity and isotropy might affect how we interpret our observations \cite{Clarkson_2007,Bolejko:2010ec,Ellis:2011hk,Secrest:2022uvx}. Indeed, the possibility that the accelerated expansion of the Universe \cite{1998AJ....116.1009R,1999ApJ...517..565P}, discovered in the 1990s, could be explained in terms of an inhomogeneous universe was extensively explored \cite[e.g.][]{Zalaletdinov:1996aj,Clarkson_2011,ben2012backreaction}. Void models \cite{Garcia-Bellido:2008vdn,Clifton:2008hv,Wiltshire2008}, Swiss-cheese models \cite{EinsteinStraus1945,Flanagan:2011tr,Bolejko:2012ue,Korzyński_2015}, and lattice ones \cite{Clifton:2009jw,Clifton:2009bp}, were all investigated, as was the possibility that small-scale inhomogeneities might react back on the expansion of space \cite{Buchert_2012}; yet, accurate predictions from these various models were found to be exceedingly difficult. Currently, it is not clear that inhomogeneities alone can explain all of the accelerated expansion of the Universe \cite[see, e.g.,][]{Bull:2011wi,Green_2014,Buchert_2015,Green_2016,Clifton_2019} although they may lead to non-negligible effects.

If the observed accelerated expansion is expected to be driven by a cosmological constant, current observations appear to deviate from this, perhaps because large-scale inhomogeneities bias our observables subtly, and make a $\Lambda$-universe mimic a phantom-like dark-energy component. Unlike genuinely evolving dark energy, inhomogeneities and their impact on light propagation are a \emph{necessary consequence} of general relativity and must exist at some level. Accurately computing their effects remains a major challenge without a fully reliable, comprehensive framework, so we take a broad approach to estimate their possible influence by considering three manifestations: (a) a lumpy universe with matter clumps separated by empty space; (b) small-scale structure that may react back on the expansion rate; and (c) mild, but sufficiently large-scale, inhomogeneities that can affect cosmological observables. Somewhat surprisingly, we find below that all three can make a $\Lambda$-dominated universe appear to contain evolving dark energy to some degree. Let us proceed to describe these models now. 
 
\paragraph{A Lumpy Universe.} The Universe is lumpy, especially on small scales, and can be approximated as a cosmic web of clusters, filaments and walls surrounding empty space \cite{1970A&A.....5...84Z}. This contrasts with homogeneous, Friedmann--Lema\^{i}tre--Robertson--Walker (FLRW) cosmology, which assumes a uniform matter density $\rho_{\rm m}$. In a homogeneous Universe, light propagates through space-time with continuously sourced Ricci curvature, $R\propto \rho_{\rm m}$, whereas in a lumpy Universe light rays spend much of their path in near-empty regions with little or no Ricci curvature, $R\simeq 0$  \cite[e.g.][]{RevModPhys.29.432,Clifton:2009jw,Kalbouneh:2025jnp}.

Cosmological observers infer the underlying cosmology by comparing measured distances to standard rulers and their redshifts with FLRW distance--redshift relations, but this can be biased because light does not propagate on FLRW geodesics \cite{Zeldovich1964}. While relativistic effects on ultra-large scales in {\it linear} large-scale structure are well studied (see e.g.\ \cite{Bonvin:2011bg,ChallinorLewis2011,Jeong:2011as}), here we focus instead on the {\it non-linear} impact of the small-scale cosmic web. In this regime, voids with $\delta \leq 0$ occupy far more volume than filaments or pancakes with $\delta>0$; so even though $\abs{\Phi}\ll 1$, its statistics become non-trivial.

The Dyer--Roeder approximation \cite{DyerRoeder1972,Dyer:1973zz} attempts to capture the fact that for most of its journey, light from faraway galaxies travels through voids, interspersed with high density regions as it crosses walls or filaments. This modifies the angular diameter distance as a function of redshift, which now satisfies the following ODE:
\begin{equation}\label{eqn:Dyer-Roeder}
    \frac{\md^2 D_A}{\md z^2} + \left[\frac{\md \ln H }{\md z} + \frac{2}{1+z}\right]\frac{\md D_A}{\md z} + \frac{3\alpha \Omega_{\rm m} H_0^2D_A}{2H(z)^2(1+z)^{-1}}  = 0\, .
\end{equation}
Here $D_A$ is the angular diameter distance, $z$ is the redshift, $H(z)$ is the Hubble parameter ($H_0\equiv H(0)\equiv h \times 100\, {\rm km~s}^{-1}~{\rm Mpc}^{-1}$ is the Hubble constant), $\Omega_{\rm m}$ is the matter energy density today, and the Dyer--Roeder parameter $\alpha$ models the amount of lumpiness (or emptiness) in the Universe, with the case $\alpha = 1$ corresponding to standard FLRW propagation. The initial conditions are $D_A(z=0) = 0$ and $\left.\mathrm{d}D_A/\mathrm{d}z\right|_{z=0} = c/H_0$.
Given that $\alpha-1$ determines the deviation from uniform spatial density, we expect $\alpha-1 \sim \langle \delta \rangle_{\rm los}$ \cite{Bolejko2011}, where $\langle \delta \rangle_{\rm los}$ is the average density fluctuation over the line of sight, whence here we use
\begin{equation}
    \alpha(z) \equiv 1+\left[\alpha_0-1\right]D_+(z)\,,
\end{equation}
where $D_+(z)$ is the linear growth factor, normalised to unity at $z=0$. We leave $\alpha_0 \equiv \alpha(0)$ as a free parameter (there are other choices of the functional form of $\alpha(z)$, e.g. see \cite{Dyer:1973zz,Bolejko2011,Dhawan_2018,Koksbang2021}). We further allow for $h$ to change with $\alpha_0$ so that the Dyer--Roeder model matches \emph{Planck}'s measurement of the angular-diameter distance to the surface of last scattering; this leads to an insignificant change, from the \emph{Planck} 2018 best-fit value $h=0.674$ \cite{Planck:2018vyg} to $h_{\alpha}$ close to it (e.g.~$h_{\alpha}=0.682$ for $\alpha_0 = 0.95$).

\paragraph{The effect of back-reaction.} 
The Einstein field equations are highly non-linear, so any split into a homogeneous, isotropic background plus inhomogeneities must be consistent. Although CMB observations and the success of linear perturbation theory show the metric is close to FLRW on relevant scales, accumulated small-scale non-linearities may react back on the expansion, yielding a small, but possibly non-negligible, correction due to the Einstein tensor’s non-linearity \cite{Ellis:1987zz,Futamase1996,Buchert:1999er,GreenWald2011,Buchert_2012,Gasperini:2011us,Rasanen2012,BentivegnaBruni2016,Goldbergetal2017,Adamek_2019,Gallagher_2020,HoltzmanMaes2024}. This can be written as an effective large-scale Einstein equation for the metric $g$ with an extra stress tensor $B_{\mu\nu}$, quadratic in small-scale metric fluctuations:\footnote{$g$ here corresponds to $g^0+\langle g^1\rangle$ of ref.~\cite{Ginat2021}.}
\begin{equation}\label{eqn:Einstein with back-reaction}
G_{\mu\nu} + \Lambda g_{\mu \nu} = 8\pi G\left(T_{\mu\nu} + B_{\mu \nu}\right)\,.
\end{equation}
If $g$ is homogeneous and isotropic, then statistically so is $B$, so we take
$
B_{\mu\nu}\equiv \rho^{\rm b}u_\mu u_\nu + P^{\rm b}\left(g_{\mu\nu}+u_\mu u_\nu\right),
$
with $u_\mu$ denoting the co-moving 4-velocity. Here $\rho^{\rm b}$ and $P^{\rm b}$ are distinct from the usual perturbative density/pressure fluctuations in $T_{\mu\nu}\equiv T^0_{\mu\nu}+\delta T_{\mu\nu}$, where $T^0_{\mu\nu}$ is exactly homogeneous and isotropic, and $\delta T_{\mu\nu}$ arises from matter density fluctuations and peculiar velocities. 

Assuming that $\norm{B} \ll \norm{T}$ (for some appropriate norm), the full large-scale space-time metric $g$ splits into 
$
    g_{\mu \nu} \equiv \tilde{g}_{\mu \nu} + \delta g_{\mu \nu} \equiv \overline{g}_{\mu \nu} + g^{\rm b}_{\mu\nu} + \delta g_{\mu\nu}\,,
$
where $\tilde{g}$ is an FLRW space-time, and $\delta g$ is the usual metric perturbation. $\tilde{g}$ is \emph{not} the metric one obtains from equation \eqref{eqn:Einstein with back-reaction} when setting $B_{\mu \nu}=0$---rather, $\overline{g}$ is; both are FLRW metrics, but with different cosmologies. Thus, an observer making experiments in a universe described by $\tilde{g}$ would infer the wrong values of the underlying matter components (comprising $T^0_{\mu\nu}$), because the observer would only be sensitive to $B_{\mu\nu} + T^0_{\mu\nu}$. Said differently, back-reaction modifies the evolution history of the Universe slightly, leading to slight deviations in the values of cosmological parameters from the underlying ones. 

Modelling the change in cosmology due to back-reaction is a daunting task. Here, we consider a toy model: a universe comprised of a three-dimensional grid, where each cell has a single dark matter halo, of mass $M$, undergoing spherical collapse. We can then calculate $\rho^{\rm b}$ and $P^{\rm b}$ and solve the Einstein equations \eqref{eqn:Einstein with back-reaction}, perturbatively to first order in $\norm{B}$. Given the perturbed metric we can find corrections to observed distance measurements as a function of redshift; we do so by employing the cosmic-ruler formalism \cite{SchmidtJeong2012,JeongSchmidt2014,JeongSchmidt2015}. This model, whose details are in the appendix, has two free parameters: $\beta_1$ which controls the amplitude of back-reaction and $\beta_2$ which dictates when the haloes virialise.   

\begin{figure*}
    \centering
    \includegraphics[width=\textwidth]{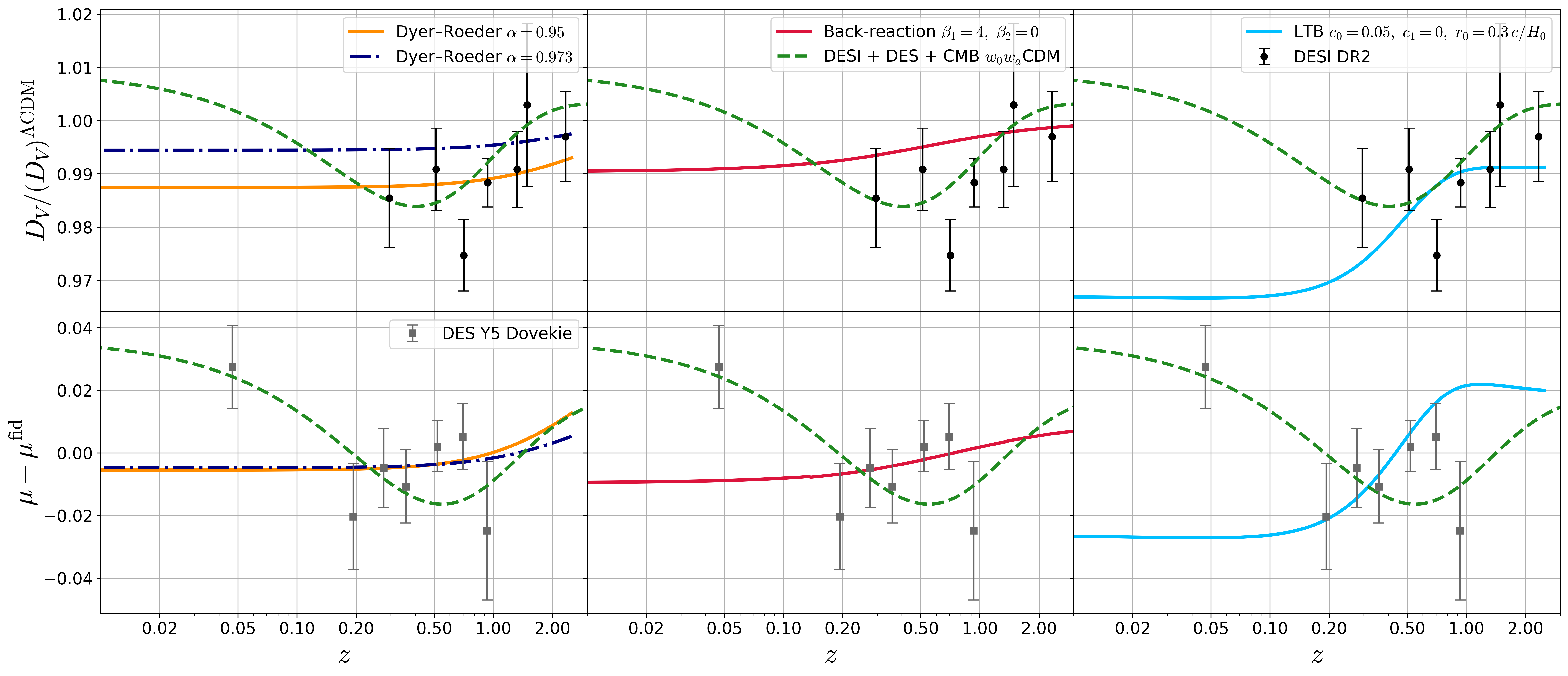}
    
    \caption{\emph{Top}: Three examples of $D_V(z)$ curves for the toy models explored here, with the DESI DR2 data-points \cite{DESI:2025zgx}, compared with the prediction of $\Lambda$CDM with \emph{Planck} \cite{Planck:2018vyg} parameters (denoted $D_V^{\rm \Lambda CDM}$). \emph{Left}: Dyer--Roeder (orange \& navy blue, dash-dotted; equation \eqref{eqn:Dyer-Roeder}), \emph{middle}: a back-reaction toy-model (red; equation \eqref{eqn:back-reaction D_V}), \emph{right}: an LTB model (light blue; equation \eqref{eqn:LTB metric}). We also show the $D_V(z)$ curve for the $w_0 w_a$CDM model preferred by DESI DR2 + DES Y5 + CMB (green, dashed) \cite[][table V, line 32]{DESI:2025zgx}; observe that it asymptotes to a value different from unity at high redshift because of the difference of the cosmological parameters from \emph{Planck}'s. \emph{Bottom}: Same as above, but for distance moduli, compared with DES Y5 (re-analysed) \cite{DES-Dovekie2026}. The distance-modulus curves were shifted to have the same mean (evaluated at the bin redshifts) as the data, with binning as in refs.~\cite{DESI:2025zgx,Lietal2026}.}
    \label{fig:example distances}
\end{figure*}

\paragraph{The Lema\^{i}tre--Tolman--Bondi model} 
Another possibility is to consider a universe with a very large-scale inhomogeneity. The work horse for such studies is the Lema\^{i}tre--Tolman--Bondi (LTB) model, which assumes the Universe is inhomogeneous but spherically symmetric \cite{Lemaitre:1933gd,Tolman:1934za,Bondi:1947fta,CRPHYS_2012__13_6-7_682_0}. The scale factor is now a function of time, $t$, and radial distance, $r$ from some preferred central point, and the metric is 
\begin{equation}\label{eqn:LTB metric}
    \mathrm{d}s^2 = -\mathrm{d}t^2 + \frac{A'(r,t)^2}{1-k(r)r^2}\mathrm{d}r^2 + A(r,t)^2\mathrm{d}\Omega^2\,.
\end{equation}
This is similar to an FLRW metric, but has a radius-dependent curvature and scale factor, $a \sim A'$. Here, ${\dot A}\equiv \partial A/\partial t$ and $A'\equiv \partial A/\partial r$. The equation that describes the expansion rate of the universe is a local version of the Friedmann equation, \emph{viz.}
\begin{equation}
    \frac{\dot{A}^2}{A^2}=\frac{F(r)}{A^3} + \frac{8 \pi G}{3}\rho_\Lambda - \frac{k(r)r^2}{A^2}\,,
\end{equation}
where $F(r)$ and $k(r)$ are free functions \cite{Garcia-Bellido:2008vdn}, and $A(r,t)$ satisfies a boundary condition $A(r,t_0) = r$ at $t=t_0$, where $t_0$ is the Universe's age. Defining $\Omega_{\rm m}$ and $\Omega_k$ in the usual way, we choose
$\Omega_{k}(r) = c_0 \mathrm{e}^{-r^2/(2r_0^2)}$ and $
\Omega_{\rm m}(r) = \Omega_{\rm m,0} + c_1\mathrm{e}^{-r^2/(2r_1^2)}$. Along a light ray, $t(z)$ and $r(z)$ satisfy the joint ODEs \cite{Garcia-Bellido:2008vdn}
\begin{align}
    \frac{\mathrm{d}t}{\mathrm{d}z} & = -\frac{A'(r,t)}{(1+z)\dot{A}'(r,t)}\,, \\ 
    \frac{\mathrm{d}r}{\mathrm{d}z} & = \frac{\sqrt{1-k(r)r^2}}{(1+z)\dot{A}'(r,t)}\,,
\end{align}
which allow to define the distance--redshift relation via $D_A(z) = A(r(z),t(z))$ along a solution of the above. 

LTB models, as alternatives to the Cosmological Principle, may be constrained in many ways (see, for example, \cite{Garcia-Bellido:2008vdn,Bull:2011wi,2011PhRvD..84b3514G,Camarena_2022}); as the main usage of these models here is illustrative, we choose to restrict ourselves to $\abs{c_0},\abs{c_1} \leq 0.1$. 

\paragraph{Results.}
We now need to model what we observe. To illustrate what we find we will discuss the observables themselves. In the case of the BAOs, these are the ``isotropised" angular diameter distances, $D_V=(cD_A^2/H)^{1/3}$, and in the case of SNe Ia, these are the apparent magnitudes, $\mu\sim \ln D_L$, all as a function of $z$. In figure \ref{fig:example distances} we plot representative curves for each of the inhomogeneous models with the BAO and SNe data. We can see that the inhomogeneous distance-curves are closer to the data than the $\Lambda$CDM Planck cosmology, albeit to different degrees, depending on the values of the model parameters. For instance, in the Dyer--Roeder case, $\alpha_0 \sim 0.95$ seems to agree well with the data. This value of $\alpha_0-1$ is somewhat lower than what one would expect for $\langle \delta \rangle_{\rm los}$ in standard $\Lambda$CDM, which is $\mathcal{O}(1\%)$ \cite{Bolejko2011,Clarksonetal2012} (although the actual range of $\alpha$ depends on the additional constraints assumed \cite{Clarksonetal2012,Dhawan_2018,Koksbang2021}), but is of the same order of magnitude. This deviation agrees with other works on ``observational back-reaction'', which found a percent level bias in distance estimates, because of light propagation in an inhomogeneous space-time \cite[e.g.][]{Macpherson:2022eve,Macpherson_2023,Macpherson_2024}. We also remark, that one expects $\alpha_0 < 1$, given that $\langle \delta \rangle_{\rm los}$ is dominated by voids; thus, curves with $\alpha_0 < 1$ should be closer to the data than those with $\alpha_0>1$ (which would lie above the $D_V/D_V^{\rm \Lambda CDM}=1$ line), as they are

\begin{figure*}
    \centering
    \includegraphics[width=\textwidth]{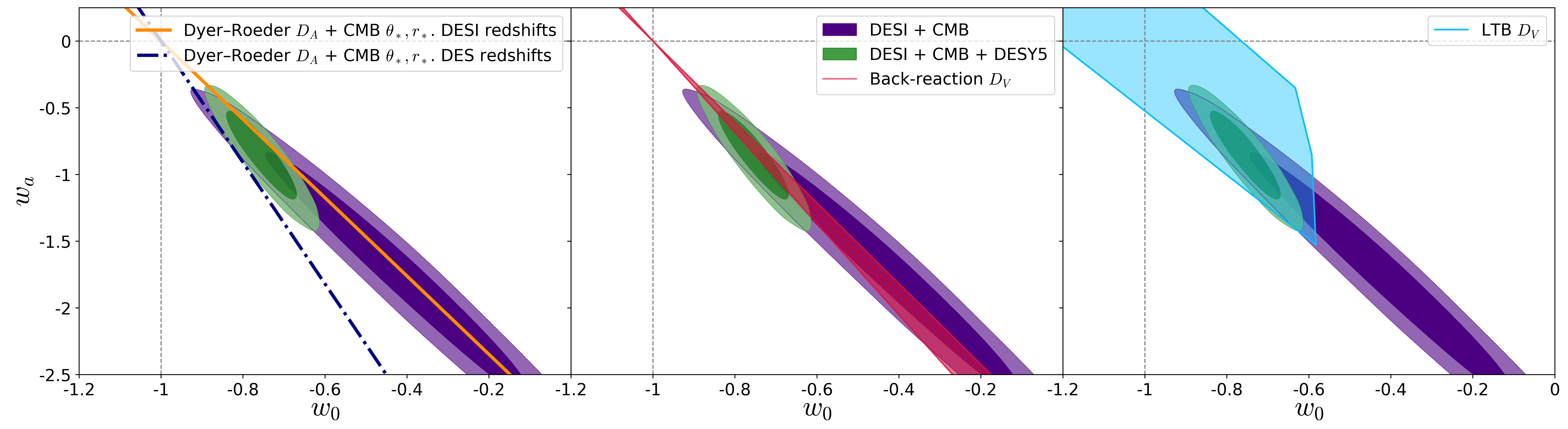}
    \caption{A plot of the $(w_0,w_a)$ values for which $D_{V,w}(z)$ lies closest to the $D_V(z)$ curve, for the three toy models considered here. \emph{Left}: Dyer--Roeder approximation (see text for details); varying $\alpha_0$ results in translation along the rays. The \emph{middle} panel displays the back-reaction toy model, where varying $\beta_1$ moves along a ray from $\Lambda$CDM ($(w_0,w_a)=(-1,0)$), while varying $\beta_2$ results in moving from one ray to another (here $\beta_2 \in [-2,5]$). \emph{Right}: the convex hull of the $(w_0,w_a)$ points for $300$ randomly-selected quadruplets of $(c_0,c_1,r_0,r_1)$ such that $\abs{c_0},\abs{c_1} \leq 0.1$. The confidence contours from ref.~\cite{DESI:2025zgx} for DESI + CMB and DESI + DES + CMB data are over-plotted.}
    \label{fig:w0-wa plane}
\end{figure*}

For the back-reaction toy model, the amplitude $\beta_1$ of $B_{\mu\nu}$ required to make a significant difference to the $D_V(z)$ curve is about half-an-order-of-magnitude larger than what is na\"{i}vely expected to be the case. That is, the $\beta_1=4$ curve, which still deviates too weakly from $\Lambda$CDM to agree with the BAO data as well as the $w_0w_a$CDM one, still has an amplitude larger than what the toy model would ordinarily predict. However, the sign of the change, where $D_V[\tilde{g}]$ is in general smaller than $D_V[\overline{g}]$, brings this curve closer to the data. We remark that here, as in the Dyer--Roeder case, the $D_V(z)/D_V^{\rm \Lambda CDM}$ curve is not versatile enough to have a dip at $z \sim 0.5$, followed by a rise, making it potentially less successful a fit than $w_0w_a$CDM. 

In the case of the LTB model, we see the characteristic slope (and depression) at low $z$ that one expects \cite{Clifton:2008hv} if one is living in a void (or a region of negative curvature) before asymptoting to the flat, $\Lambda$ dominated region of the space time, probed at higher redshift. As above, this model predicts monotonically changing distance with redshift, unlike $w_0w_a$CDM.

While looking directly at the observables is instructive, one can use the effective equation of state to highlight the behaviour which seems to indicate that the data deviate from a $\Lambda$-dominated universe. 
The standard assumption is that the expansion rate satisfies the standard Friedmann equation
$
    H^2=H_0^2\left[\Omega_{\rm m} a^{-3}+(1-\Omega_{\rm m})S(a)\right]\, , 
$
where we have introduced a free function of the scale factor, $S(a)$. Note that there are other dependencies (such as radiation, massive neutrinos, \emph{etc}., that we have ignored). We can further assume that
$
    S(a)=\exp\left\{-3\int_a^1\left[1+w(a')\right]\mathrm{d}a'\right\}
$
where we have introduced a free function of the scale factor, $w(a)$ which we will call the equation of state. It is often convenient to approximate $w(a)=w_0+w_a(1-a)$ where $w_0$ and $w_a$ \cite{Chevallier:2000qy,Linder:2002et}. Thus, given measurements of the expansion rate of the Universe, we can constrain $H_0$, $\Omega_{\rm m}$ and crucially, $w_0$ and $w_a$, which encode the behaviour of the dark-energy component.

To approximate how we measure the expansion rate of the Universe, we generate mock data (specifically $D_V$ and $\mu$) for each of our inhomogeneous models, in the redshifts probed by the BAO and SNe Ia data. We then use the associated uncertainties to construct a mock likelihood which we can then use to infer the best-fit parameter from the FLRW model, with the aforementioned $S(a)$. That is, we find the $w_0w_a$CDM $D_{V}(z)$ and $\mu(z)$ curves which are closest to the inhomogeneous $D_V(z)$ and $\mu(z)$ ones, i.e.~we associate a point $(w_0,w_a)$ with each inhomogeneous model---with each value of $\alpha_0$ in the Dyer--Roeder case, pair $(\beta_1,\beta_2)$ in the back-reaction case, and $(c_0,c_1,r_0,r_1)$ in the LTB case (see appendix for details).  
A far more accurate approach has been extensively developed in \cite{Garcia-Garcia:2019cvr,Traykova:2021hbr,Wolf:2023uno,Shlivko:2024llw,Wolf:2024eph,Wolf:2025jed}. While not a substitute for such a full analysis, our approach does find, for each of the inhomogeneous models we are considering, the predicted locus on the $(w_0,w_a)$ plane corresponding to it, where we can compare with the predictions from the actual data, which have already been fitted with a realistic likelihood and covariance matrix to the BAO/SNe data. 

The result of doing so is presented in figure \ref{fig:w0-wa plane}. The first striking feature is that {\it all} toy models, for the parameters ranges we consider, lie in the $\set{w_0> -1,w_a<0}$ quadrant, and overlap with the inferred confidence contours from the DESI and DES Y5 data. This is the case for the $\alpha_0 \leq 1$ Dyer--Roeder models and the $\beta_1 \geq 0$ back-reaction models, in agreement with ref.~\cite{Dhawan_2018}, who found a degeneracy between $\alpha$ and $w$ from SNe data. For the LTB model, the inferred values of $w_0$ and $w_a$ also cover the DES and DESI data contours but sweep out a much larger swathe. Furthermore, the further away from a $\Lambda$-dominated universe these models are, the larger these parameters have to be. Thus, the toy models in figure \ref{fig:example distances} suggest that inhomogeneities might furnish a systematic for inferring properties of evolving dark energy.

The left panel of figure \ref{fig:w0-wa plane} displays the Dyer--Roeder curves that arises from comparing $D_{A,w}$ to the Dyer--Roeder $D_A(z)$. The Dyer--Roeder equation only evolves $D_A(z)$ and assumes an unmodified $D_H = c/H(z)$ and we keep it as such in the comparison. We plot the $(w_0,w_a)$ curves from comparing at DES redshifts only (blue, dash-dotted) and DESI ones (orange); the joint one is halfway between the two.


\paragraph{Discussion.}
An inhomogeneous, $\Lambda$-dominated Universe may look like one with evolving dark energy. We have presented a first, rough attempt at modelling the effect of three distinct types of inhomogeneities, and have shown that the resulting, effective, equation of state is remarkably similar to what is obtained from measurements of BAO by DESI and SNe Ia by DES, thus alleviating the tension with $\Lambda$CDM. The magnitude of the deviation of the effective $w_0$-$w_a$ curve from a cosmological constant, due to inhomogeneities, seems to be too small to explain the \emph{entirety} of the signal; but this should be determined by future, more detailed, analyses. The hints, seen here, that inhomogeneities might act as a source of systematic error for dark-energy inference, therefore warrant further study. In particular, they will exacerbate the problem of un-determination of dark energy \cite{Wolf:2023uno}.

The way forward is not easy, as accurately modelling large-scale inhomogeneities is a notoriously hard problem---hence our approach here. There have been a number of attempts at quantifying the effects, either through perturbation theory, ray tracing through $N$-body simulations in FLRW background, in idealised inhomogeneous relativistic models or even in full numerical relativity simulations. But, again, these are useful to get a rough (but not precise) estimate of the magnitude of the effect that inhomogeneities might have on distance indicators. For quantitative inference, what is required is an accurate, but relatively flexible model that can be incorporated into statistical analysis pipelines, used to constrain the expansion of the universe. Only then will we be able to truly infer the fundamental physics behind the expansion of the Universe.

\paragraph{Acknowledgements.} We thank K.~Bolejko, T.~Clifton, C.~Clarkson, H.~Desmond, G.~Ellis, R.~Ewart, C.~Garcia-Garcia, Z.~Haiman, B.~Kocsis, R.~Maartens, M.~Nastac, A.A.~Schekochihin and W.~Wolf for many valuable conversations. This work was supported by the STFC (grant No.~ST/W000903/1), and by a Leverhulme Trust International Professorship Grant (No.~LIP-2020-014). YBG's work was partly supported by the Simons Foundation via a Simons Investigator Award to A.A.~Schekochihin. PGF is also supported by the Beecroft Trust. 

\bibliographystyle{apsrev}
\bibliography{refs}
\onecolumngrid
\appendix

\section{A model for back-reaction}
Here, we will satisfy ourselves with considering a toy model which uses the halo model: it can be shown that in general, $P^{\rm b} = (2K+U)/V_{\rm small~scale}$, i.e.~twice the kinetic energy density, plus the potential energy density of the small scales \cite{Baumannetal2012,Ginat2021}. We now envisage a universe comprised of a three-dimensional grid, where each cell has a single halo, of mass $M$, undergoing spherical collapse. Each grid cell contributes $(2K+U)_{\rm cell}/V_{\rm cell}$; averaging over a volume enclosing many cells, and assuming that $M$ is distributed according to a halo mass-function $\phi(M) = \mathrm{d}n/\mathrm{d}M$, yields
\begin{equation}
    P^{\rm b}(t_0) = \int \mathrm{d}M \int_{t_0}^{\infty} \mathrm{d}t ~B_{\rm cell}(t_0;t,M) \frac{\partial \phi(M)}{\partial t} \, ,
\end{equation}
where $B_{\rm cell}(t0;t,M)$ is $(2K+U)$ from a spherical-collapse model of the halo of mass $M$, for which the maximum expansion time is $t_{\max} = t/2$. The term $\partial\phi/\partial t$ denotes the rate at which haloes of mass $M$ form (per unit volume). Note that one need not integrate over $t \leq t_0$ because a halo which virialises (i.e.~forms) at $t$ ceases to contribute. We use a \citet{Tinkeretal2008} mass-function for $\phi(M,z)$. This yields $P^{\rm b}(t_0)$ for every $t_0$. 

We then solve the Einstein equations \eqref{eqn:Einstein with back-reaction}, perturbatively to first order in $\norm{B}$. The solution, $\tilde{g}$, may be written as 
\begin{equation}
    \mathrm{d}\tilde{s}^2 = a^2\left(-\left[1+2\Phi_{\rm b}(\eta)\right]\mathrm{d}\eta^2 + \left[1-2\Phi_{\rm b}(\eta)\right]\delta_{ij}\mathrm{d}x^i\mathrm{d}x^j\right)\,, 
\end{equation}
where the potential $\Phi_{\rm b}$ describes the deviation of the true expansion history from the background one, from $a(\eta)$ corresponding to $\overline{g}$. The unknowns $\Phi^{\rm b}$ and $\rho^{\rm b}$ satisfy the two ODEs 
\begin{align}
    & \Phi_{\rm b}'' +3\mathcal{H}\Phi_{\rm b}'+3H_0^2\Omega_{\Lambda}a^2\Phi_{\rm b} = 4\pi G a^2 P^{\rm b}\,, \\ & 
    4\pi G a^2 \rho^{\rm b} = -3\mathcal{H}\left(\Phi_{\rm b}' + \mathcal{H}\Phi_{\rm b}\right)\, ,
\end{align}
where $'=\partial/\partial \eta$ and $\mathcal{H} = a'/a$, which are readily solved.\footnote{These equations, together with the Friedmann equations for $a(\eta)$, are equivalent to solving the Friedmann equations for $\tilde{g}$ (at the appropriate order in $\norm{B}$). We find it more instructive to view them perturbatively, though.} 
Then, given $\Phi_{\rm b}$, it is straightforward to calculate the corrections to observed distance measurements as a function of redshift; we do so by employing the cosmic-ruler formalism \cite{SchmidtJeong2012,JeongSchmidt2014,JeongSchmidt2015}. The relevant rulers are $\mathcal{M}$ and $\mathcal{C}$, which respectively quantify distortions transverse to the line of sight and along it. For our special case, expressions for these rulers are, to first order in $\norm{B}$: 
\begin{align}
    \mathcal{C} & = \frac{\mathrm{d}\ln r_0}{\mathrm{d}\ln \tilde{a}}\mathcal{T} - \Delta \ln a \left[1-H(z)\frac{\mathrm{d}}{\mathrm{d}z}\left(\frac{1+z}{H(z)}\right)\right] - \Phi_{\rm b} +2\frac{1+z}{H(z)}\Phi_{\rm b}' \, , \\ 
    \mathcal{M} & = 2\frac{\mathrm{d}\ln r_0}{\mathrm{d}\ln \tilde{a}}\mathcal{T} + 2\Delta \ln a\left[\frac{(1+z)}{H\tilde{\chi}} - 1\right] + 2\Phi_{\rm b}-\frac{4}{\tilde{\chi}}\int_0^{\tilde{\chi}}\Phi_{\rm b}\mathrm{d} \chi + \frac{2}{\tilde{\chi}}\int_0^{t_0}\Phi_{\rm b}\mathrm{d}t\, ,
\end{align}
where we used $\Delta \ln a = -H_0 \int_0^{t_0} \Phi_{\rm b}\mathrm{d}t$, 
\begin{equation}
    \mathcal{T} = H(z) \int_0^{\eta}\Phi_{\rm b} a(\eta')\mathrm{d}\eta' - H_0\int_0^{\eta_0}\Phi_{\rm b}a(\eta')\mathrm{d}\eta' -\Phi_{\rm b}(\eta_0) + \Phi_{\rm b}\,,
\end{equation}
and for the BAO feature, $\mathrm{d}\ln r_0/\mathrm{d}\ln \tilde{a} = 1$ \cite{JeongSchmidt2014}.
In terms of these rulers, 
\begin{align}
    \frac{D_A[\tilde{g}]}{D_A[\overline{g}]} & = 1-\frac{\mathcal{M}}{2}\, , \\ 
    \frac{D_V[\tilde{g}]}{D_V[\overline{g}]} & = \left(1-\frac{\mathcal{M}}{2}\right)^{2/3} \left(1-\mathcal{C}\right)^{1/3}\, .\label{eqn:back-reaction D_V}
\end{align}

We elevate this toy model into a two-parameter family of such models, by inserting two free parameters: (i) we re-scale $B_{\rm cell} \mapsto \beta_1 B_{\rm cell}$, making the amplitude of back-reaction a free parameter; (ii) in computing $B_{\rm cell}$, we terminate the spherical collapse of the halo slightly before $t = 2t_{\max}$, because spherical collapse predicts infinite densities at that time, and hence infinite $B_{\rm cell}$; therefore, there is some freedom in choosing when to deem the halo `virialised'. We introduce a parameter $\beta_2$ to do that, by deeming the halo virialised when $2K+U = -\beta_2 (K+U)$ for the first time after $t>t_{\max}/2$. The na\"{i}ve values for the parameters are thus $\beta_1 = 1$, $\beta_2 = 0$, but we consider $\beta_2 \in [-2,5]$ and any $\beta_1$, as an arbitrary choice.  

\section{Mock Likelihood}
In order to associate a point $(w_0,w_a)$ to our inhomogeneous toy models in figure \ref{fig:w0-wa plane}, we compute a distance measure from the model, for an array of parameter values. For each value of the model parameter ($\alpha_0$ in the case of Dyer--Roeder, $\beta_{1,2}$ for back-reaction, and $c_{0,1}$, $r_{0,1}$ for LTB) we compare the curves $D_V^{\rm model}(z)$ and $\mu^{\rm model}(z)$ with the FLRW ones with evolving dark energy, denoted $D_{V,w}(z)$ and $\mu_w(z)$. For each model-parameter value, we minimise the following quantity over $(w_0,w_a)$
\begin{equation}
    \chi^2 \equiv p_{\textrm{SNe}}\sum_{i \in \textrm{SNe bins}} \left[\frac{D^{\textrm{model}}_A(z_i) - D_{A,w}(z_i)}{\delta \mu(z_i) \,D_{A,w}(z_i)}\right]^2 + p_{\rm BAO}\sum_{i \in \textrm{BAO bins}} \left[\frac{D^{\textrm{model}}_V(z_i) - D_{V,w}(z_i)}{\delta D_V(z_i)}\right]^2\,,
\end{equation}
where $\delta D_V(z_i)$ is the error of the experimental measurement of $D_V(z_i)$, and similarly for $\delta \mu(z_i)$, and $p_{\textrm{SNe}}$ and $p_{\rm BAO}$ are weights adjusted so that each data-set has a $50\%$ contribution to $\chi$. This associates a point $(w_0,w_a)$ to each inhomogeneous model, where the distance--redshift curve is closest to the evolving dark-energy one, at the observed redshifts. 

For example, for the Dyer--Roeder model, this process finds a curve $(w_0(\alpha_0),w_a(\alpha_0))$, where $(w_0(1),w_a(1)) = (-1,0)$, corresponding to $\Lambda$CDM. For $\alpha_0$ close to $1$, by Taylor-expanding the solution, one would have 
\begin{align}
    w_0(\alpha_0) & = -1 + m_0(\alpha_0-1) + \mathcal{O}(\alpha_0-1)^2 \,, \\ 
    w_a(\alpha_0) & = m_a(\alpha_0-1) + \mathcal{O}(\alpha_0-1)^2 \,,
\end{align}
where $m_0$ and $m_a$ are constants. Substituting the second equation into the first yields a linear relation between the $w_0(\alpha_0)$ and $w_a(\alpha_0)$, of the form $w_a = (m_a/m_0)(w_0+1)$, which is indeed what we see in figure \ref{fig:w0-wa plane}. A similar argument also applies to the back-reaction model, for a fixed $\beta_2$, (by $\beta_1$ playing the role of $\alpha_0$).

\end{document}